# FLEXIBLE AUDIO STREAMS

**ABSTRACT**

Tremendous research effort was invested in audio browsers and machine learning techniques to decode the structure of Web pages in order to put them into an audio format. In this paper, we address a simpler and efficient solution for the creation of an audio browser of VOICEXML generated from RSS/Atom stream feeds. We developed a multimodal (audio and graphical) portal application that offers RSS/Atom feeds. By utilizing sing our system, the user can interact using voice or graphic commands, listen and watch digital content, such as news, blogs feeds, podcasts, and even access email and personal schedules. The portal system permits the use of security credentials (user/password authentication) to collect secure RSS/Atom stream in the multimodal browser to connect the user to specific personal services. A series experiments have been conducted to evaluate the performance of the RSS reader and navigator. Our system is extremely beneficial for a wide range of applications, from interfaces for the visual impaired users to browsers for mobile telephonic interfaces.

## 1. INTRODUCTION

Screen readers are motivated by the fact that blind users cannot quickly identify relevant information in Web pages. Most audio browsers existent today (e.g. GW Micro Window Eyes [GW Micro, 2006], IBM Home Page Reader [IBM, 2003], Freedom Scientific Jaws [Freedom Scientific, 2006], the systems from the W3c Alternative Web Browsing initiative [W3C, 2005]) are simple sequential screen readers. Few advanced audio browsers use either shortcut commands or machine learning to identify sections in Web pages and the most probable interaction workflows. Even these systems are very annoying and difficult to use by the visual impaired users. In this paper, we present our multimodal RSS feed browser that improves the interaction speed and the user experience.

Our browser provides a solution for aggregating content and applications from various RSS feeds. It has an integrated audio and graphical interface and a single signed-on (SSO) approach for security. This was achieved by using generic JSP and XSLT interfaces that transform RSS feeds into VoiceXML audio format.

We also provide the possibility of using secure RSS feeds by providing security credentials (user/password authentication) at the connection to a personalized feed (e.g. Google Calendar Atom format [Google Calendar, 2006]). The portal supports a variety of technologies: RSS 2.0, Atom 0.3, SSL/HTTPS, and HTTP authentication and transforms the feeds in VoiceXML version2.0.

We carried out  a set of relevant experiments in which we obtained results (average access time and satisfaction rates) significantly better than by using the standard audio browsers. For details please see the results section.

## 2. METHODS

The structure of our RSS multimodal reader and navigator portal is depicted in Figure 1. The portal application server collects the RSS/Atom stream feeds and uses a JSP and XSLT converter to transform these feeds into VoiceXML or XHTML+VoiceXML dialog menus (see Figure 2). The user interacts with the list of titles in the feed and can select a specific link to follow using mixed-initiative interaction. The user can access the RSS feeds in two different ways. First, one can use a multimodal browser (e.g. the audio components from Opera Internet browser)which uses XHTML+VoiceXML and  a synchronized dual voice and graphical interface. Second, the user can access just the voice interface using a VoiceXML browser (e.g. IBM ViaVoice software).

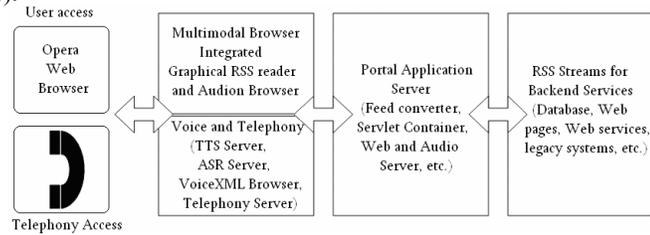

Figure 1. Voice Portal Interface to RSS/Atom Streams

The portal offers the possibility of including shortcuts to different operations, allowing the personalization of the user experience. The users interact with the browser through a dialog manager that uses a collection of dialog templates to convert an RSS feed into appropriate VoiceXML dialog. This dialog manager can apply different transformation patterns for various RSS feeds, and can distinguish between reading and writing operations following the Google Calendar API protocol.

The components of the RSS feeds (i.e. the titles, the descriptions and the links to the following pages) are extracted from the RSS source and a tree is created with all possible dialog routes. The words are extracted from the collected titles and are searched in a list of possible shortcuts while a ranking weight can be given to certain entries based on these words. This way, the user can search for a certain subset of entries in the RSS feed by only pronouncing shortcut words. The grammar is automatically generated from the title of the entries in the RSS feed.

```
<?xml version="1.0" encoding="UTF-8"?>                                    RSS feed
<rss version="2.0">
<channel>
    <title>Google News</title>
    <link>http://news.google.com/?ned=iePkh8BM9EOKzgxVohwEbP1vMBELmPv-4xLLNcfZPrVNqUhvrAF9KD4I</link>
    <language>en</language>
    <pubDate>Fri, 23 Feb 2007</pubDate>
    ...
    <item>
        <title>US seeks more stringent UN sanctions against Iran - San Jose Mercury News</title>
        <link>http://news.google.com/news/url?sa=T&ct=us/0-0-0&fd=R&</link>
        <pubDate>Fri, 23 Feb 2007</pubDate>
        <description>The United States will seek tougher UN sanctions ...</description>
    <item>
    <item>
        <title>Senator takes a meeting at CAA - The Southern</title>
        <link>http://news.google.com/news/url?sa=T&ct=us/0-0-0&fd=R&</link>
        <pubDate>Fri, 23 Feb 2007</pubDate>
        <description>Continuing the week's intriguing fusion of ...</description>
    <item>
    ...
<?xml version="1.0" encoding="iso-8859-1"?>
<!DOCTYPE html PUBLIC "-//VoiceXML Forum//DTD XHTML+Voice 1.2//EN"
    "http://www.voicexml.org/specs/multimodal/x+v/12/dtd/xhtml+voice12.dtd">
<html xmlns="http://www.w3.org/1999/xhtml"                              XHTML+VoiceXML translation
    xmlns:vxml="http://www.w3.org/2001/vxml"
    xmlns:ev="http://www.w3.org/2001/xml-events"
    xmlns:xv="http://www.voicexml.org/2002/xhtml+voice"
    xml:lang="en-US">
    <head>
    ...
    <vxml:form id="rss_form">
        <vxml:field name="rss_items" xv:id="rss_items_name" modal="true">
            <vxml:grammar>
                <![CDATA[
                #JSGF V1.0 iso-8859-1;
                grammar items;
                public <items> = <NULL> { $= new Array; } (<item> [and] { $.push($item) } )+;
                <item>  = US seeks more ($="1")  |
                          Senator takes a meeting ($="2")  |
                          ... ;
                ]]>
            </vxml:grammar>
            <vxml:prompt bargein="true"> Please say the items. </vxml:prompt>
        </vxml:field>
    </vxml:form>
    <xv:sync xv:field="#rss_items_name" xv:input="items"/>
    ...
    </head>
    <body>
    ...
    <select name="items" id="items" multiple="multiple" size="10" width="100%">
        <option> US seeks more stringent UN sanctions against Iran - San Jose Mercury News </option>
        <option> Senator takes a meeting at CAA - The Southern </option>
    </select>
    ...
    </body>
</html>
```

Figure 2. RSS to XHTML+VoiceXML translation

The system can keep a dialog context by using the shortcut words and weights propagated in the dialog interaction. This way, in following dialogs with the same user, the system can keep a ranking of possible paths in dialog and assign new shortcuts to these paths. All previous dialog paths and their ranking are kept in history, so the user can access the most relevant dialog paths very easily.

## 3. RESULTS

A series experiments have been conducted to evaluate our RSS voice navigator's accuracy and quantitative performance against, the state-of-the-art screen reader GW Micro Window Eyes, currently one of the most common used screen readers. Other screen readers delivered the same set of characteristics with this product. In our evaluation we used twenty RSS feeds and their corresponding Web sites (spanning four content domains: news, blogs, e-mail, and calendar). First, we had simple cases in which the user was dealing with transformed news and blogs RSS feeds into XHTML+VoiceXML feeds. Second, we had a more complex dialog for consulting a schedule using the Google Calendar RSS feed or for reading Google e-mail. The important points to consider are: we used Web applications that have also a Web page and a RSS feed, and, in all experiments, the users could see the RSS feed, but the interaction was realized only with audio commands.

We realized that audio browsing with screen readers is very time-consuming. Our program does most of what screen readers do because currently most websites offer RSS feeds. Compared with GW Micro Window Eyes, our system achieved browsing time reduction of 23%. Because our program does not need filtering through irrelevant information, we can rapidly identify a short list of relevant possible actions. Little content analysis was necessary for our browser since the description already gives us a summarization of the content.

We compared our program with Window Eyes using two different metrics. First, we measured the average time to select the desired information from one page or to complete one successful atomic step of interaction (considering the task completion success rate as 100%) (see Figure 3). Second, we measured the quality cost, as the users' perception of the system. The user satisfaction rating was computed by having users complete a survey at the end of the experiment (see Figure 4). In all the experiments we obtained results significantly better with our browser than with the standard audio browser.

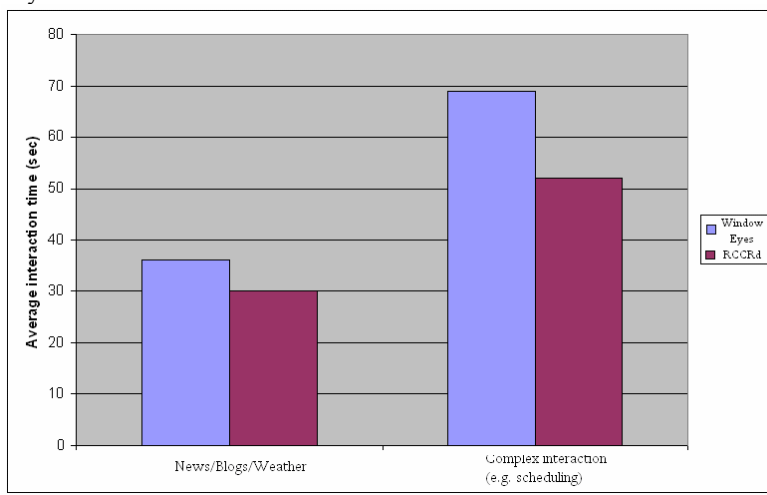

Figure 3. Average time to select the desired information from one page or to complete one successful atomic step of interaction

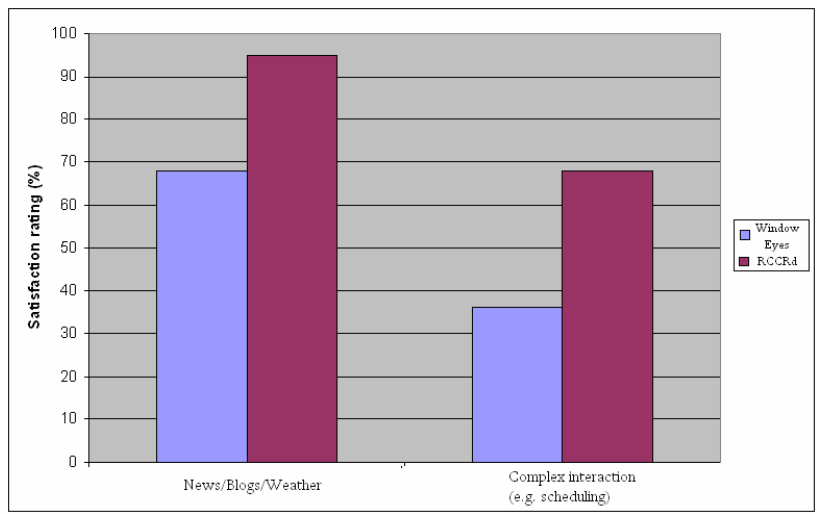

Figure 4. User satisfaction

## 4. CONCLUSION

In this paper, we have described the design and implementation of the RSS Reader and Navigator, a voice RSS non-visual Web browsing system. We presented experimental evidence of the efficiency of RSS-directed Web browsing using audio. The following are a few potentially useful areas for further research. Machine learning techniques can be employed in ranking algorithms for complex dialog management features for using the dialog context and previous dialog paths in predicting the future user's behavior in browsing.